\documentclass[12pt]{article}
\usepackage{latexsym,graphicx,multirow}
\usepackage{float}
\usepackage{amssymb}
\usepackage{amsmath}
\usepackage{amscd}
\usepackage{amsthm}
\usepackage[left=2cm,top=2.5cm,right=2.5cm,bottom=1.5cm]{geometry}
\usepackage{hyperref}
\usepackage{epstopdf}
\usepackage{cite}
\usepackage{color}
\usepackage[utf8]{inputenc}
\usepackage[T1]{fontenc}
\begin{document}
	
	\begin{center}
		\large{\bf{ R$\grave{e}$nyi Holographic Dark Energy models in Teleparallel gravity}} \\
		\vspace{10mm}
		\normalsize{ Vinod Kumar Bhardwaj$^1$, Archana Dixit$^2$, Anirudh Pradhan$^3$, Symala Krishannair$^{4}$   }\\
		\vspace{5mm}
		\normalsize{$^{1,2,3}$Department of Mathematics, Institute of Applied Sciences and Humanities, G L A University\\
			Mathura-281 406, Uttar Pradesh, India}\\
			\vspace{5mm}
		\normalsize{$^{4}$Department of Mathematical Sciences, Faculty of Science, Agriculture and Engineering,University of 
		Zululand, Kwadlangezwa 3886, South Africa} \\
		\vspace{2mm}
		$^1$E-mail:  dr.vinodbhardwaj@gmail.com\\
		\vspace{2mm}
		$^2$E-mail: archana.ibs.maths@gmail.com\\
		\vspace{2mm}
		$^3$E-mail: pradhan.anirudh@gmail.com \\
		\vspace{2mm}
              $^{4}$E-mail:krishnannairs@unizulu.ac.za\\

		\vspace{10mm}
		
	\end{center}
	\begin{abstract}
		In this paper, we have investigated the physical behavior of cosmological models in the framework of modified Teleparallel gravity. 
		This model is established using a Renyi holographic ``dark energy model (RHDE) with a Hubble cutoff. Here we have considered a 
		homogeneous and isotropic Friedman universe filled with perfect `fluid. The physical parameters are derived for the present model 
		in Compliances with 43 observational Hubble data sets (OHD). The equation of state (EoS) parameter in terms of $H(z)$ describes a 
		the transition of the universe between phantom and non-phantom phases in the context of $f(T)$ gravity. Our model shows the violation of 
		strong energy condition (SEC) and the weak energy condition (WEC) over the accelerated phantom regime. We also observed that these 
		models occupy freezing regions through $\omega_{D} -\omega_{D}^{'}$ plane. Consequently, our Renyi HDE model is 
		supported to the consequences of general relativity in the framework of $f(T)$ modified gravity.

	\end{abstract}
	
	

\section{Introduction}
According to the numerous observations \cite{ref1,ref2,ref3,ref4}, the accelerated expansion of the Universe is due to the presence of an exotic kind of
energy, ``called dark energy (DE)". The nature and the cosmological origin of DE are still enigmatic. To describe the phenomenon of DE,
several models have been presented. According to several findings, DE should behave like a fluid with ``negative pressure, counterbalancing
the action of gravity, and speeding up the universe"\cite{ref5,ref6,ref7}. The general methodology is to define the dynamics of the universe
by assuming the source of DE represented as a non-zero ``cosmological constant $\Lambda$”, connected to ``vacuum quantum field
fluctuations”\cite{ref8,ref9,ref10}.\\
   
   As proposed by observations, the universe is homogeneous, flat, isotropic, and conforms to FLRW space-time. The existing state of the universe
    encouraged scientists further to investigate Einstein's general relativity (GR) to discover the reasons.
    A cosmological constant is added in GR field equations representing the mysterious form of energy. But it has several drawbacks, like 
    the fine-tuning and the coincidence problems, and its effects are only seen at ``cosmological sizes "rather than `Planck scales' \cite{ref11}.
    As a result, various modified theories of gravity have been introduced in the literature to overcome the problems with GR. In the framework of
    modified gravity, it is feasible to change the ``Ricci scalar (R)" in the `Einstein-Hilbert action' by arbitrary functions of $R$  
    \cite{ref12,ref13,ref14,ref15}. However, namely the ``Teleparallel Equivalent of General Relativity (TEGR)"  is proposed in the literature
    \cite{ref16,ref17,ref18}. \\

 In TEGR, $T$ (torsion scalar) represents the gravitational action. The $f(T )$ gravity is developed as the function of $T$ and  extension of 
 TEGR \cite{ref19,ref20,ref21}. The estimations  of GR agree to coincide with TEGR, while the similar doesn't occur for $f(T )$  with respect 
 to $f(R)$ gravity \cite{ref21}. This motivates researchers to propose the various cosmological modifications in $ f(T)$ gravity 
 \cite{ref22}-\cite{ref31}. Among these, the generalized `teleparallel theory of gravity' has recently attracted much attention as a possible 
 explanation for DE. In this simplification the Lagrangian of teleparallel gravity, the torsion scalar $T$, is changed by its generic function f(T) 
 \cite{ref27}. The f(T) gravity models utilize the Weitzenb'ock connection, which acquires the torsion  and is responsible for the accelerating 
 expansion of the universe \cite{ref32}. This idea was first proposed by Einstein in 1928 under the name `Fern-Parallelismus or teleparallelism' 
 \cite{ref33,ref34}.\\
 
The second-order field equations of the $f(T)$ theory have a major advantage over the fourth-order field equations of the f(R) theory 
\cite{ref35,ref36,ref37}. Authors \cite{ref38} have discussed the detected universe's acceleration without DE and performed observational 
viability tests for several $f(T)$ models using SNIa data. Using the Bianchi type I (BI) universe, Sharif and Shamaila \cite{ref39} developed 
certain $f(T)$ models. They have also calculated the EoS parameter for `two teleparallel models'. Yang developed three new $f(T)$ models and 
explained their scientific consequences and cosmic behavior \cite{ref40}. Aly \cite{ref41} investigated the $f(T)$ gravity in the framework 
of the THDE. In the same context, Ayman and  Selim worked on the behavior of the fractal cosmology in $f(T)$ gravity\cite{ref42}. 
Cosmology of $f(T )$ gravity in a ``holographic dark energy and non isotropic" background  discussed in \cite{ref43}.\\

In $f(T )$  cosmology, Capozziello {\it et al.} \cite{ref44} present a ``model-independent approach for numerically solving the modified 
Friedmann equations". They discussed the behavior of several cosmological parameters with redshift in this hypothesis. In recent studies 
\cite{ref45,ref46,ref47}, HDE model has been considered broadly and analyzed as  $\rho\propto\Lambda^{4}$  using the connections between IR, cut off 
and entropy such that $\Lambda^{3}L^{3}\leq S^{3/4}$. In a similar context, the energy density of HDE model is obtained by the relation of IR cut-offs 
and `Bekenstein-Hawking entropy $S = A/4G$'. The energy density of vacuum depends on the Ricci scalar UV cut-off (particle, event, and Hubble horizons) 
of the universe and the infrared (IR) cut-off.\\

A lot of literature is available on the investigations of different IR cut-off \cite{ref48,ref49,ref50,ref51}. In past few years, various HDE models 
have been investigated for the cosmological and gravitational incidences using different entropies such as Tsallis \cite{ref52,ref53,ref54}, 
Reyni \cite{ref55,ref56} and Sharma-Mittal \cite{ref57}. Chen \cite{ref58} introduced recent developments on holographic entanglement entropy.
Among these models, a new dark energy model proposed by Moradpour {\it et al.} \cite {ref56} named the Rényi holographic dark energy (RHDE) model for 
the cosmological and gravitational investigations shows more stability by itself. Several researchers have discussed RHDE in different theories of gravity. \\
	
Keeping in view the above motivations, we have investigated the Reyni HDE using the FLRW metric in the context of $f(T)$ modified gravity.
The manuscript is arranged in the following manner: In Sect. $2$, the basic equations in $f (T )$ gravity are derived with the framework of the FLRW metric.
Renyi's holographic dark energy model is discussed in Sect. $3$. In Sect. $4$, we have discussed the solutions of the field equations with observational
constraints. Sect. $5$ is devoted to the investigation of the energy conditions. In Sect. $6$, we have obtained the $\omega_{D}-\omega_{D}^{'}$ plane.
Sect. $7$ is devoted to the conclusions.

		
\section{ Basic Field Equations in f(T) Gravity}

 The orthonormal tetrad components $e_{A}x^{\mu} $are used to study the teleparallel gravity. Where an index $A$  and $B$ runs over 0, 1, 2, 3 for 
 the tangent space at each point $x^{\mu}$ of the manifold. Their relation to the metric $g_{\mu v}$ is given by

\begin{equation}
	\label{1}
	g_{\mu\nu}= \eta A B e_{\mu}^{A}e_{\mu}^{B}
\end{equation}
where $\mu$ and $\nu$ are coordinate indices on the manifold run over 0, 1, 2, 3, and $e_{\mu}^{A}$ forms the tangent vector of the manifold.\\

The `torsion $T^{\rho}_{\mu\nu}$ and contorsion $K^{\mu\nu}_{\rho}$ tensors' are read by

\begin{equation}
	\label{2}
T^{\rho}_{\mu\nu}\equiv e_{A}^{\rho}(\partial_{\mu}e_{\nu}^{A}-\partial_{\mu}e_{\mu}^{A})
\end{equation}

\begin{equation}
	\label{3}
	K_{\rho}^{\mu\nu}\equiv \frac{-1}{2}(T^{\mu \nu}_{\rho}-T^{\nu \mu}_{\rho}-T^{\mu \nu}_{\rho})
\end{equation}

In GR, the Lagrangian density is characterized by the Ricci scalar $R$', whereas the teleparallel Lagrangian density is described by 
the ~torsion scalar $T$', which is defined as
\begin{equation}
	\label{4}
	T\equiv S_{\rho}^{\mu\nu} T^{\rho}_{\mu \nu}
\end{equation}
where
\begin{equation}
	\label{5}
	 S_{\rho}^{\mu\nu}\equiv\frac{1}{2}( K_{\rho}^{\mu\nu}+\delta^{\mu}_{\rho}T^{\alpha \nu}_{\alpha}-\delta^{\nu}_{\rho}T^{\alpha \mu}_{\alpha})
\end{equation}
Subsequently, the modified teleparallel action for f(T) theory is specified by \cite{ref33}
\begin{equation}
	\label{6}
	I=\frac{1}{16\pi G}\int(d^{4} x |e| (T+f(T))
\end{equation}
where $T$ is the torsion and $T=-6H^2$ and $f(T)$ is a differentiable function of the torsion scalar $T$, 
$|e|$ = $det (e^{A}_{\mu})$ = $\sqrt{-g}$ and consider $c = 16\pi G=1$ . We notices that in the action (\ref{6}), we have omitted any matter 
distribution. For more detail derivation giving a clear picture of the relation to relativity.\\

The space-time metric of Friedmann-Lemaitre-Robertson-Walker (FLRW) is provided by
\begin{equation}
\label{7}
ds^2=-dt^2+a(t)^2 dx^2
\end{equation}
 where $ a(t)$  is the average scale factor. Using variation, the modified Friedmann equations in $ f(T) $ gravity are written as \cite{ref26}: 
	
\begin{equation}
\label{8}
H^2 =\frac{1}{3} \rho_{m} -\frac{1}{6} f-2 H^2 f_{T} 
\end{equation}
\begin{equation}
\label{9}
	(H^2)' =\frac{2 p_{m}+6 H^2+12 H^2 f_{T}+f}{24 H^2 f_{TT}-2f_{T}-2} 
\end{equation}	
where, prime represents the derivative with respect to $ \ln a $, $ f_{T}=\frac{df}{dT} $, $ f_{TT}=\frac{d^2 f}{dT^2} $, $ \rho_{m} $ is energy 
density of dark matter and $ p_{m} $ is the pressure of dark matter. 
The modified Friedmann equations in general relativity are given as:
\begin{equation}
	\label{10}
	H^2 =\frac{1}{3}\left(\rho_{m}+\rho_{D}\right) 
\end{equation}
\begin{equation}
\label{11}
	(H^2)' =-\left(\rho_{m}+p_{m}+\rho_{D}+p_{D}\right) 
\end{equation}
For a pressure less dark matter, $ p_{m}=0 $. On comparing the field Eq.(\ref{8})-Eq.(\ref{9})  with Eq.(\ref{10})-Eq.(\ref{11}), the expression 
for the energy density $ \rho_{D} $ and pressure $ p_{D} $ are given as \cite{ref26}: 

\begin{equation}
\label{12}
\rho_{D} =\frac{1}{2} \left(-f+2 T f_{T}\right) 
\end{equation}
\begin{equation}
\label{13}
p_{D}=\frac{f-T f_{T}+2 T^2 f_{TT}}{1+f_{T}+2 T f_{TT}}
\end{equation}
For our $f(T)$ gravity model, this is the pressure and dark energy density.


\section{ Renyi Holographic Dark Energy Model}

We have consider a $k$ states structure having  probability distribution $P_{i}$ and fulfill the condition $\sum_{i=1}^{k} P_{i}=1$. 
Renyi entropy is a recognized generalized entropy parameter \cite{ref59}.
\begin{equation}
	\label{14}
	S_{R}=\frac{1}{\delta}ln\sum_{i}^{k}P_{i}^{1-\delta}~~~~~ S_{T}=\frac{1}{\delta}ln\sum_{i=1}^{k}P_{i}^{1-\delta}- P_{i},
\end{equation}
where  $U\to$  real parameter, $\delta=1-U $. By using Eq. (\ref{14}), now we have obtain the relation,

\begin{equation}
\label{15}
S_{R}=\frac{1}{\delta}ln(1+\delta S_{T}).
\end{equation}

In Eq. (\ref{15}) the Bekenstein entropy is 
$S_{T} = \frac{A_{r}}{4}$, ~ and $A_{r}= 4\pi L^{2}$.
This gives the Renyi entropy of the system as $S_{R}=\frac{1}{\delta}ln(1+\delta\pi L^{2})$ \cite{ref56,ref60} .

In this model we have taken the Renyi holographic dark energy (RHDE) written as \cite{ref56},
\begin{equation}
\label{16}
\rho_{_D}=\frac{3 c^2}{8 \pi G L^2}\left(1+\pi \delta L^2\right)^{-1}.
\end{equation}
 The holographic principle (HP) states that the DE density is proportionate to the square of the Hubble parameter, i.e., $\rho_{D} \propto H^{2}$ 
 \cite{ref61}. According to this principle, the specified choice can resolve the fine-tuning problem. Here the
Renyi entropy $ S_{R}=\frac{1}{\delta}\ln\left(\frac{\delta}{4}A_{r}+1\right) =\frac{1}{\delta}\ln\left(\pi \delta L^2+1\right)$, 
where $ A_{r}=4\pi L^2 $ and $ L =H^{-1}$ is IR cut-off. 

The Renyi HDE is obtained as

\begin{equation}
	\label{17}
	\rho_{_D}=\frac{3 c^2 }{8 \pi G}\frac{H^2}{\left(1+\frac{\pi \delta}{H^2} \right)}.
\end{equation} 

\section{Solutions of the Field Equations with Observational Constraints}
The power-law scale factor is the suitable choice for explaining the matter radiation, and DE-dominated universe. So we choose the scalar of 
the form Sharma {\it et al.} \cite{ref62} to describe the possible supper accelerated transition.

\begin{equation}
\label{18}
a=\left(n D t\right)^{\frac{1}{n}}
\end{equation}
where $ n $ and $ D $ are positive nonzero constants. From Eq. (\ref{18}), the Hubble parameter obtained as $H=\frac{\dot a}{a}=\frac{1}{nt}$. 
The following relationship connects the scale factor ($ a $) and the redshift ($ z $).
\begin{equation}
\label{19}
a=  \frac{a_{0}}{(1+z)}
\end{equation}
As a result, the Hubble parameter in terms of redshift read as:	
\begin{equation}
\label{20}
H(z)=H_{0} (1+z)^{n}
\end{equation}
where $ H_0\to $  present value of Hubble parameter.\\

In this section, we identify constraints on the model parameters $H_{0}$ and $n$ bounded by the model under the recent $43$ observational Hubble data 
(OHD) in the range of $0\leq z \leq 1.965$.  The data of all 43 H(z) points are compiled in table I of \cite{ref63,ref64}.
The reason for using this data is that OHD data produced the cosmic chronometric (CC) technique is model-independent. As detected by observational 
Hubble data (OHD), the current value of the Hubble constant is $H_{0} = 69.45\pm 1.8$  $km/s/Mpc$ and $n = 0.7907\pm0.0687$. We defined $\chi^{2}$ 
for constraining model parameters  $H_{0}$ and $n$ with given as

\begin{equation}\label{21}
\chi^{2}\left(H_{0}, n\right)=\sum_{i=1}^{43} {\frac{\left(H_{th}(i)-H_{ob}(i)\right)^2}{\sigma(i)^2}}
\end{equation}
where $H_{th}(i)$ are the theoretical values of $H(z)$ as per (\ref{20}) and $\sigma_{i}'s$ are errors in the observed values of H(z). 
The 1- Dimensional marginalized distribution and 2- Dimensional contours with 68.3 $\%$, 95.4 $\%$ and 99.7$\%$ confidence level respectively 
are obtained for our model as depicted in Fig. $1$.
 
\begin{figure}[H]
	\centering
	\includegraphics[scale=1]{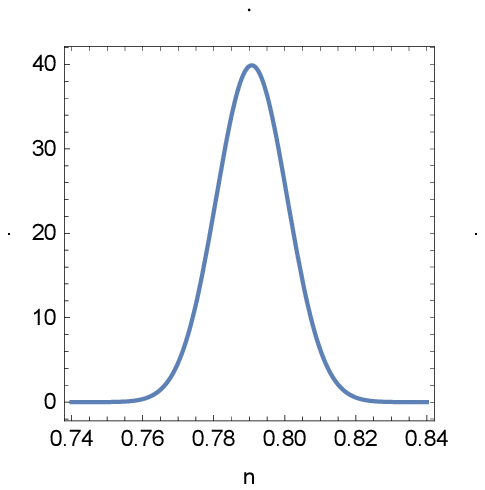}\\
	\includegraphics[scale=0.5]{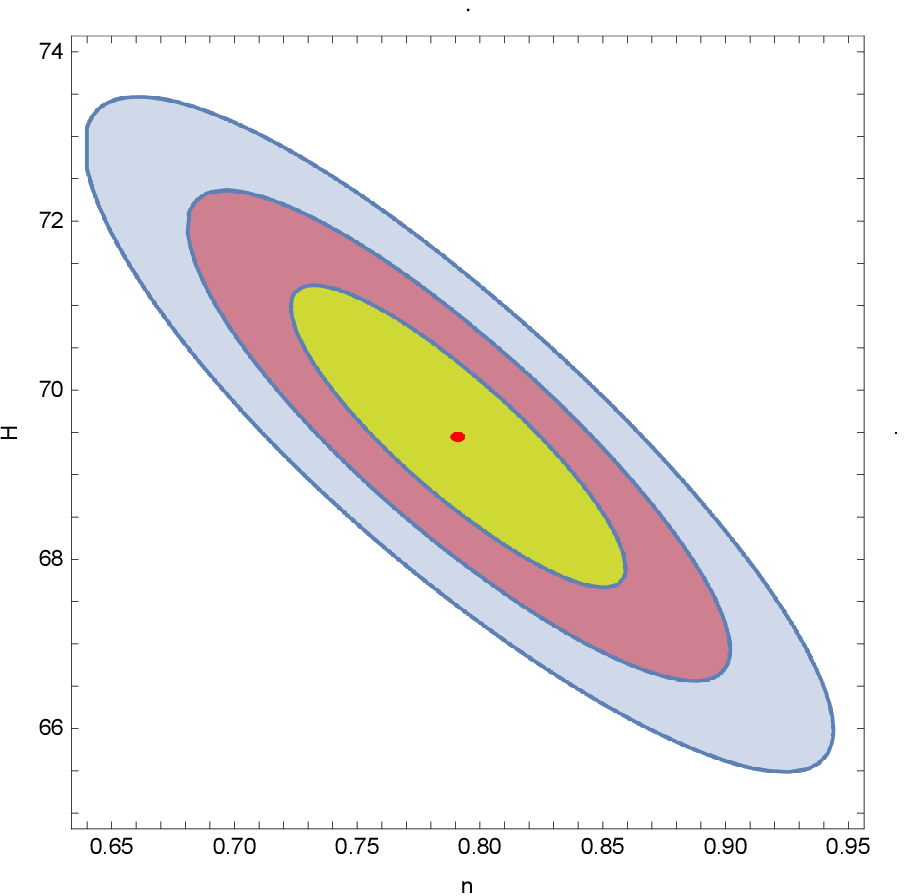}
	\includegraphics[scale=1]{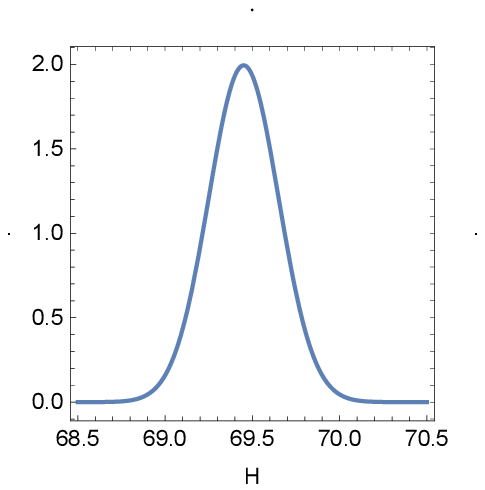}
	\caption{1-Dimensional marginalized distribution and 2-Dimensional contour plots
		 with best fitted values as $n = 0.7907\pm0.0687$ \& $H = 69.45\pm 1.8$ in the $n - H$ plane.}
		 \label{fig1}
\end{figure}
\begin{figure}[H]
	\centering
	\includegraphics[scale=0.8]{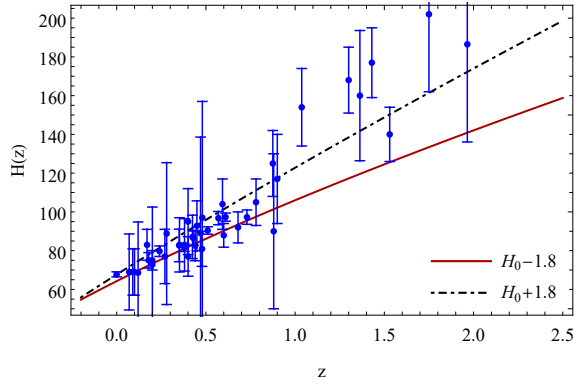}
	\caption{Error bar plots of the Hubble data set}\label{fig2}
\end{figure}
The values of Hubble constant at different redshifts have been
estimated by many cosmologists \cite{ref62,ref63,ref65} by using the `galaxy clustering method and differential age approach'. They determined several
observed values of Hubble constant (Hob)  along with corrections in the range $0 \leq z \leq 1.965$ (Table-1). The observed and theoretical values
are found to agree quite well. In the figure-$2$, the dots in this diagram indicate the 43 observed Hubble constant (Hob) values with corrections.
The linear curves show the theoretical values of the  Hubble constant $(H)$  with marginal corrections.

\begin{table}[H]
	\caption{\small The behaviour of Hubble parameter $H(z)$ with redshift}
	\begin{center}
		\begin{tabular}{|c|c|c|c|c|c|c|c|c|c|}
			\hline
			\tiny	$S.No$  &	\tiny  $Z$ & \tiny $H (Obs)$ & \tiny $\sigma_{i}$ & \tiny References & \tiny $S.No$  &	\tiny  $Z$ & \tiny $H (Obs)$ & \tiny $\sigma_{i}$ & \tiny References \\
			\hline
			\tiny	1	& \tiny 0	  &\tiny 67.77 & \tiny 1.30 & \tiny\cite{ref66} & \tiny	24	& \tiny 0.4783	 &\tiny 80.9  & \tiny9 & \tiny \cite{ref74}   \\
			
			\tiny	2	& \tiny 0.07  &\tiny 69    & \tiny 19.6 & \tiny \cite{ref67} & \tiny	25	& \tiny 0.48	 &\tiny 97 & \tiny60 & \tiny \cite{ref69}  \\

			\tiny	3	& \tiny 0.09	 &\tiny 69  & \tiny 12 & \tiny \cite{ref68}   & 	\tiny	26	& \tiny 0.51	 &\tiny 90.4  & \tiny1.9 & \tiny\cite{ref73} \\  
			\tiny	4	& \tiny 0.01	 &\tiny 69  & \tiny 12 & \tiny \cite{ref69}   & \tiny	27	& \tiny 0.57	 &\tiny 96.8  & \tiny3.4& \tiny \cite{ref77}  \\	
			\tiny	5	& \tiny 0.12	 &\tiny 68.6  & \tiny26.2 & \tiny \cite{ref67}  & \tiny	28	& \tiny 0.593	 &\tiny 104 & \tiny 13 & \tiny \cite{ref70}  \\ 
			%
			\tiny	6	& \tiny 0.17	 &\tiny 83  & \tiny 8 & \tiny \cite{ref69}   & \tiny	29	& \tiny 0.60	 &\tiny 87.9  & \tiny6.1 & \tiny\cite{ref75} \\ 
			%
			\tiny	7	& \tiny 0.179	 &\tiny 75  & \tiny 4  & \tiny \cite{ref70}  & \tiny	30	& \tiny 0.61	 &\tiny 97.3  & \tiny2.1 & \tiny\cite{ref73} \\ 		
			\tiny	8	& \tiny 0.1993	 &\tiny 75  & \tiny 5  & \tiny\cite{ref70}   & \tiny	31	& \tiny 0.68	 &\tiny 92  & \tiny 8 & \tiny\cite{ref70}   \\ 
			\tiny	9	& \tiny 0.2	 &\tiny 72.9  & \tiny 29.6  & \tiny \cite{ref67} & 	\tiny	32	& \tiny 0.73	 &\tiny 97.3 & \tiny 7 & \tiny \cite{ref75}   \\ 
			\tiny	10	& \tiny 0.24	 &\tiny 79.7  & \tiny 2.7 & \tiny \cite{ref71} & \tiny	33	& \tiny 0.781	 &\tiny 105 & \tiny 12 & \tiny \cite{ref70}   \\		
			\tiny	11	& \tiny 0.27	 &\tiny 77  & \tiny 14 & \tiny \cite{ref69}   & \tiny	34	& \tiny 0.875	 &\tiny 125  & \tiny 17 & \tiny \cite{ref70}  \\ 
			\tiny	12	& \tiny 0.28	 &\tiny 88.8  & \tiny 36.6 & \tiny \cite{ref67}  & 	\tiny	35	& \tiny 0.88	 &\tiny 90  & \tiny 40 & \tiny  \cite{ref69} \\
			%
			\tiny	13	& \tiny 0.35	 &\tiny 82.7 & \tiny 8.4 & \tiny \cite{ref72}    & \tiny	36	& \tiny 0.9	 &\tiny 117  & \tiny 23 & \tiny  \cite{ref69} \\ 
			\tiny	14	& \tiny 0.352	 &\tiny 83 & \tiny 14 & \tiny \cite{ref70}    & \tiny	37	& \tiny 1.037	 &\tiny 154  & \tiny 20 & \tiny \cite{ref70} \\ 
			\tiny	15	& \tiny 0.38	 &\tiny 81.5  & \tiny 1.9 & \tiny \cite{ref73}  & 	\tiny	38 	& \tiny 1.3	 &\tiny 168 & \tiny 17 & \tiny \cite{ref69} \\ 
			%
			\tiny	16	& \tiny 0.3802	 &\tiny 83 & \tiny 13.5 & \tiny\cite{ref74}  & 	\tiny	39	& \tiny 1.363	 &\tiny 160  & \tiny 33.6 & \tiny \cite{ref78} \\ 
			%
			\tiny	17	& \tiny 0.4	 &\tiny 95  & \tiny 17 & \tiny \cite{ref68}   & \tiny	40	& \tiny 1.43	 &\tiny 177  & \tiny 18 & \tiny  \cite{ref69} \\ 
			%
			\tiny	18	& \tiny 0.4004	 &\tiny 77 & \tiny10.2 & \tiny\cite{ref74}   & 	\tiny	41	& \tiny 1.53	 &\tiny 140  & \tiny  14 & \tiny \cite{ref69} \\ 
			\tiny	19	& \tiny 0.4247	 &\tiny 87.1  & \tiny 11.2 & \tiny\cite{ref74}  & 	\tiny	42	& \tiny 1.75	 &\tiny 202  & \tiny 40 & \tiny \cite{ref69}\\ 
			\tiny	20	& \tiny 0.43	 &\tiny 86.5  & \tiny 3.7 & \tiny\cite{ref71}  & \tiny	43	& \tiny 1.965	 &\tiny 186.5  & \tiny 50.4 & \tiny \cite{ref78}  \\ 
			\tiny	21	& \tiny 0.44	 &\tiny 82.6  & \tiny 7.8 & \tiny\cite{ref75}  & \tiny		& \tiny 	 &\tiny  & \tiny  & \tiny  \\ 
			\tiny	22	& \tiny 0.44497	 &\tiny 92.8  & \tiny 12.9 & \tiny\cite{ref74}  & 	\tiny		& \tiny 	 &\tiny   & \tiny   & \tiny   \\
			%
			\tiny	23	& \tiny 0.47	 &\tiny 89 & \tiny49.6  & \tiny\cite{ref76}    & 	\tiny		& \tiny     	 &\tiny    & \tiny      & \tiny   \\
			\hline	
		\end{tabular}
	\end{center}
\end{table}

\begin{equation}\label{22}
\rho_{D}=\frac{6 c^2 H_0^4 (z+1)^{4 n}}{\pi  \delta +H_0^2 (z+1)^{2 n}}
\end{equation}

Solving Eqs. (\ref{22}), (\ref{9}) and (\ref{12}), we get:

\begin{eqnarray}\label{23}
f(T)&=&-6 \sqrt{-H_0^2 (z+1)^{2 n}} \bigg[2 \sqrt{\pi } c^2 \sqrt{\delta } \tanh ^{-1}\left(\frac{\sqrt{-H_0^2}}{\sqrt{\pi } 
\sqrt{\delta }}\right)-2 \sqrt{\pi } c^2 \sqrt{\delta} \nonumber\\
&\times&\tanh ^{-1}\left(\frac{\sqrt{-H_0^2 (z+1)^{2 n}}}{\sqrt{\pi } \sqrt{\delta }}\right)+2 c^2 \sqrt{-H_0^2 (z+1)^{2 n}}-(1+2 c^2) 
\sqrt{-H_0^2}\bigg]
\end{eqnarray}
Here Eq. (\ref{23}) represents the $f(T)$ function for proposed model. From Eq. (\ref{13}), the pressure for the model determined as:

\begin{equation}\label{24}
p_{D}=\frac{12 \pi  c^2 \delta  H_0^4 (z+1)^{4 n}}{-2 \pi  \left(2 c^2-1\right) \delta  H_0^2 (z+1)^{2 n}+\left(1-2 c^2\right) H_0^4 (z+1)^{4 n}+
\pi ^2 \delta ^2}
\end{equation}

By using  Eqs. (\ref{22}) and (\ref{24}) the equation-of-state parameter $\omega_{D}$ is given by
	
\begin{equation}\label{25}
\omega_{D}=\frac{\pi  \delta  \left(6 \pi  \delta +6 H_0^2 (z+1)^{2 n}\right)}{-6 \pi  \left(2 c^2-1\right) \delta  H_0^2 (z+1)^{2 n}+
3 \left(1-2 c^2\right) H_0^4 (z+1)^{4 n}+3 \pi ^2 \delta ^2}
\end{equation}
	
The dark energy density parameter $\Omega_{D}$ is given by 
	
\begin{equation}
\label{26}
\Omega_{D}=\frac{c^2 H_0^2 (z+1)^{2 n}}{\pi  \delta +H_0^2 (z+1)^{2 n}}
\end{equation}
We can obtain $\omega_{D}^{'}$ and $\Omega_{D}^{'}$ by differentiating  Eqs. (\ref{25}) and (\ref{26}) with respect to $\log a$ :
\begin{eqnarray}\label{27}
\omega^{'}_{D}&=&4 \pi  \delta  H_0^2 n (-z-1) (z+1)^{2 n-1} \bigg[\pi ^2 \left(4 c^2-1\right) \delta ^2\nonumber\\
&+&\left(2 c^2-1\right) H_0^2 (z+1)^{2 n}\bigg(2\pi \delta + H_0^2 (z+1)^{2 n}\bigg)\bigg]/\bigg[2 \pi \left(2 c^2-1\right) 
\delta  H_0^2 (z+1)^{2 n}\nonumber\\
&+& \left(2 c^2-1\right) H_0^4 (z+1)^{4 n}-\pi ^2 \delta ^2\bigg]^2
\end{eqnarray}
\begin{equation}\label{28}
\Omega^{'}_{D}= -\frac{2 \pi  c^2 \delta  H_0^2 n (z+1)^{2 n}}{\left(\pi  \delta +H_0^2 (z+1)^{2 n}\right){}^2}
\end{equation}
\begin{figure}[H]
	\centering
	\includegraphics[scale=0.8]{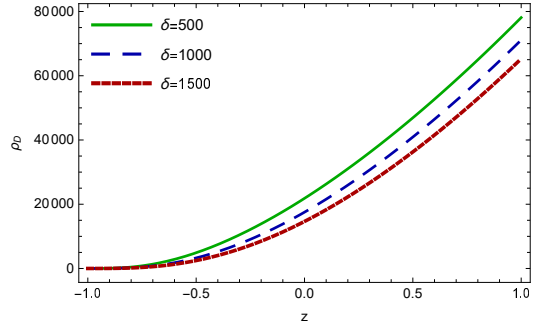}
	\caption{Variation of DE $\rho_{D}$ versus $ z $ }\label{fig3}
\end{figure}

The evolution of dark energy density with redshift is depicted in Figure $3$ for various values of $\delta$, we can see that dark energy density 
($\rho_{D}$) is a positive decreasing function of redshift throughout the evolution of the universe at the present epoch.
 
\begin{figure}[H]
	\centering
	\includegraphics[scale=0.9]{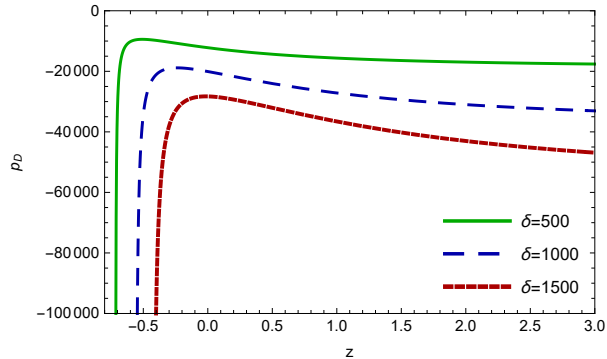}
	\caption{Variation of  $p_{D}$ versus $ z $ }\label{fig4}
\end{figure}

We have noticed in Figure 4 that the pressure decreases as the redshift increases. The pressure remains negative through the evolutionary era 
and approaches zero in the high redshift area. It has been hypothesized that the cosmic acceleration is caused by negative pressure as confirmed 
by recent estimations.

\begin{figure}[H]
	\centering
	\includegraphics[scale=0.9]{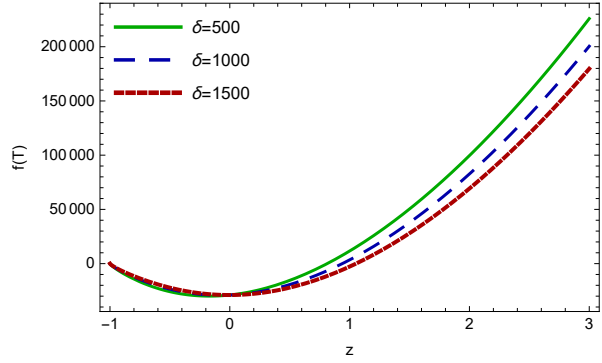}
	\caption{Variation of  $f(T)$ versus $ z $ }\label{fig5}
\end{figure}	
The behavior of $f(T)$ versus redshift is shown in Figure $5$. It is observed that $f(T)$ has an expanding behavior over the specified range. 
Various values of $\delta$, start with a high negative value and gradually approach a positive value.

\begin{figure}[H]
	\centering
	\includegraphics[scale=1]{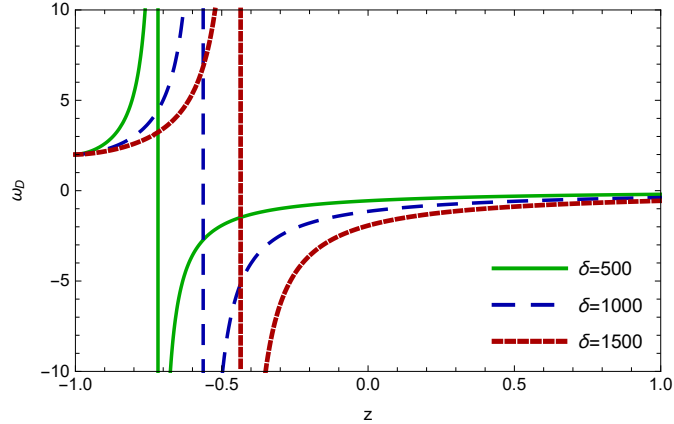}
	\caption{Variation of  $\omega_{D}$ versus $ z $ }\label{fig6}
\end{figure}
In Figure 6, we observed the dynamics of the EoS parameter against redshift ($z$) for various esteems of $\delta$ = 500, 1000, 1500. 
The EoS parameter classifies the expansion of the universe. The values of EoS parameter $-1<\omega_{D}\leq0$,  $\omega_{D}=-1 $ and $\omega_{D}<-1$ 
represents the quintessence , $\Lambda$CDM and Phantom eras respectively. In the derived model, EoS parameter $\omega_{D}$ starts with 
`quintessence era', crosses the phantom divide line, and enters in the phantom era \cite{ref79,ref80,ref81}. 
The graph predicted that the universe is under the influence of $\Lambda CDM$. It is well known that for an accelerating universe, we should have 
$\omega_{D}<-1/3$. In the later time, it shows the `stiff fluid' when $\omega_{D}=1$, the matter-dominated phase when $\omega_{D}=0$ and $\omega_{D}=1/3$ 
represents the radiation dominated phase \cite{ref19,ref82}.The graph demonstrates that dark energy is influencing the universe, as the equation of 
state predicts an accelerated expansion phase. It's worth noting that this type of crossover behavior is consistent with current cosmic observational 
findings. 

\begin{figure}[H]
	\centering
	\includegraphics[scale=0.9]{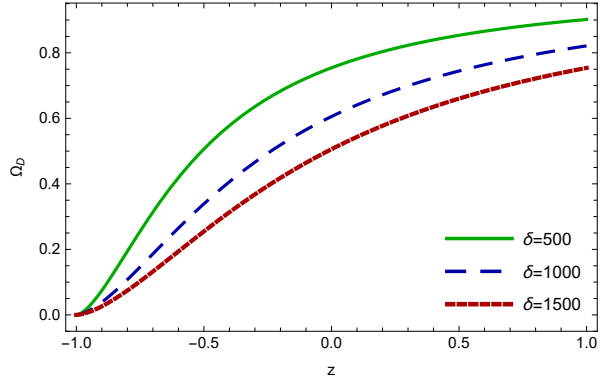}
	\caption{Variation of  $\Omega_{D}$ versus $ z $ }\label{fig7}
\end{figure}
The  density parameter ($\Omega_{D}$) for the RHDE models are depicted in  Figure 7 for different values of $\delta$ \cite{ref83,ref84}. 
Dark energy dominates the late universe and gradually evolves to $\Omega$ = 1 in the future. Here  $\Omega$ is the total density parameter. 
It is worth noting that the density parameter is decreasing in a good direction \cite{ref83,ref84,ref85}. 


\section{Energy Conditions}

The energy conditions (ECs) are significant techniques for explaining the Universe's geodesics. If we are working with a perfect fluid matter 
distribution the energy conditions recovered from conventional GR. 

There are various types of energy conditions:
\begin{itemize}
\item Null Energy Condition     (NEC): $\rho \geq 0$
\item Weak Energy Condition     (WEC): $\rho +p \geq 0 $
\item Dominant Energy Condition (DEC): $\rho-p \geq 0$
\item Strong Energy Condition   (SEC): $\rho + 3p \geq 0$.
\end{itemize}
All of the above expression for energy conditions are depend on the model parameters.

\begin{figure}[H]
	\centering
	\includegraphics[scale=0.8]{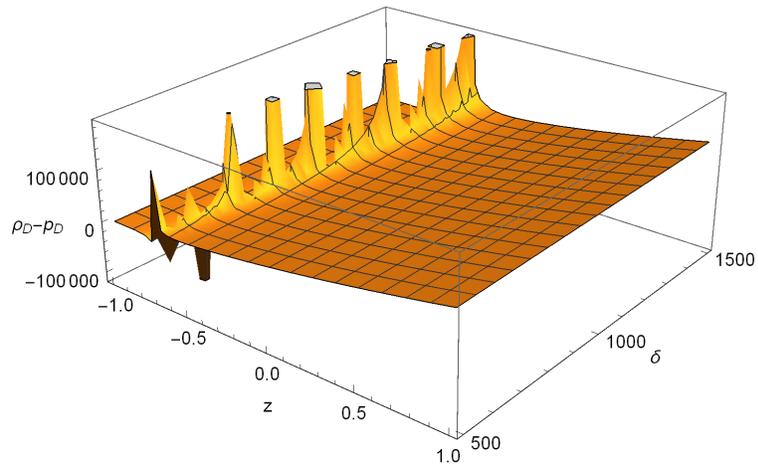}
	\includegraphics[scale=0.8]{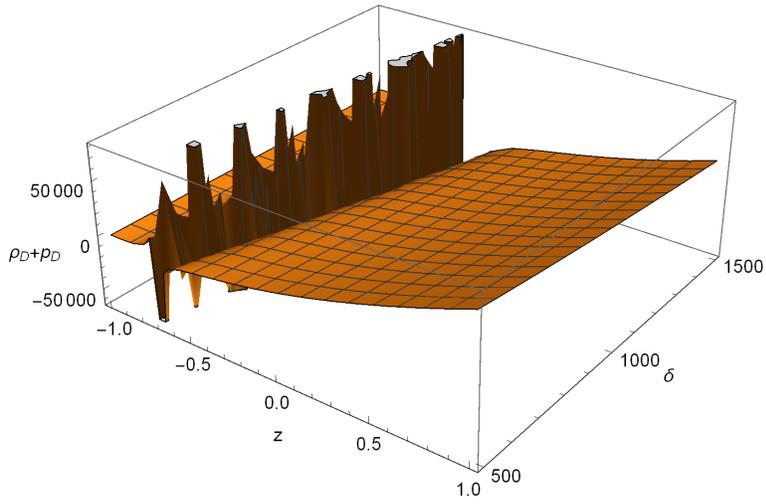}
	\includegraphics[scale=0.8]{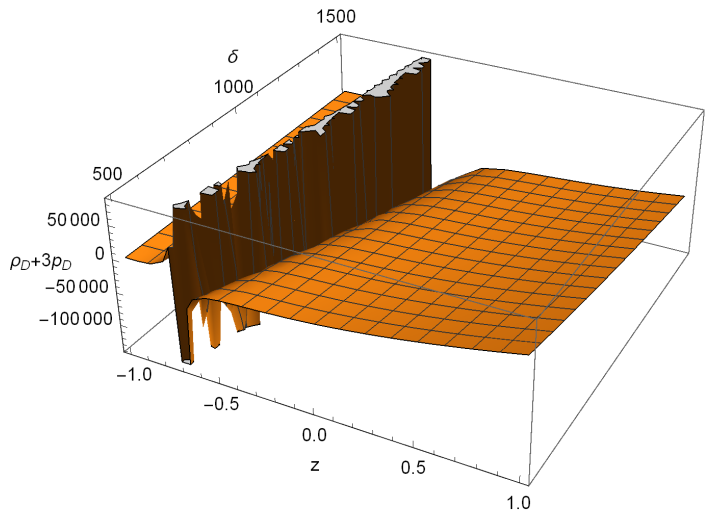}
	\caption{Variation of  $\omega_{D}^{'}$ versus $ z $ }\label{fig8}
\end{figure}
Figure 8 shows the plots of energy conditions vs. redshift. The violation of SEC is generally due to the anti-gravity matter such 
as dark energy present in the universe. For normal matter, all the energy conditions including SEC must be validated. In the derived model, 
Fig. 8 shows the validation and violation of energy conditions for the particular choice of free parameters and satisfied the energy conditions.
The WEC and DEC are satisfied initially, but both violate at the present epoch. But it is observed that the SEC does 
not satisfy $f(T)$ gravity. In the fact, these energy conditions are crucial in comprehending cosmic scenarios \cite{ref86,ref87}. 
The Hawking-Penrose singularity theorems are primarily studied using WEC and SEC. The SEC is critical to comprehending the allure of gravity 
\cite{ref88,ref89}. Now the violation of SEC, indicates clearly the expansion of the universe with acceleration which proves the correctness of 
the derived model. \\


\section{$\omega_{D}-\omega_{D}^{'}$  plane}
The authors \cite{ref90,ref91} were developed the  $\omega_{D}-\omega_{D}^{'}$  plane to examine the cosmic evolution of the quintessence 
dark energy concept. Figure 9 illustrates the phase plane dynamics of the models with  $\delta = 500, 1000, 1500$ and have freezing  like models 
$\omega_{D}\geq-1$. We have observed that our RHDE model lies  freezing region ($\omega_{D}<0$ , $\omega_{D}^{'}<0$ ).	

\begin{figure}[H]
		\centering
		\includegraphics[scale=0.9]{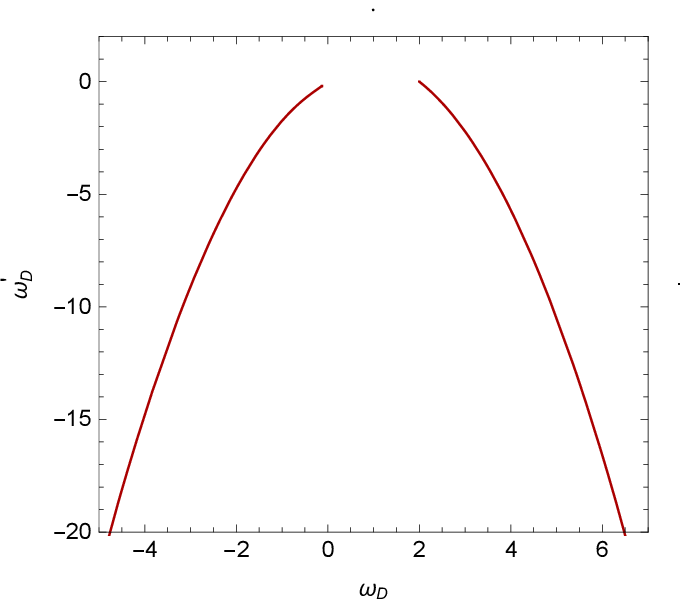}
		\caption{Plot of $ \omega_{D}-\omega'_{D} $. }\label{fig9}	
\end{figure}

It has been found that the universe's cosmic expansion accelerates more rapidly in the freezing area.
In the context of $f(T)$ modified gravity, the graphical behavior displayed ($\omega_{D}^{'}<0$ ) suggested that the Renyi HDE model is in 
the freezing region and cosmic expansion are more accelerated.


\section{Conclusions}
Using $46$ OHD points, we have calculated the cosmological parameter values of the resultant models in the present paper.
Energy density, pressure, the $f(T)$ function, the equation of state parameter, and the $\omega_{D}-\omega_{D}^{'}$ cosmic plane have all been examined.
In terms of redshift, we have obtained all these cosmological parameters here.
It is demonstrated in the current research that these results are compatible with the observation data.
The following are the model's key characteristics: 

\begin{itemize}
\item 
1-Dimensional marginalized distribution and 2-Dimensional contour plots
and Error bar plots of the Hubble data set are shown in Figs. $1$ $\&$ $2$.
We obtained throughout the evaluation, that the energy density is positive and the pressure is negative, as seen in Figs. $3$ $\&$ $4$. The energy density 
must be positive, and the pressure must be negative for cosmic acceleration. The evolution of pressure and energy density of our model satisfies WEC. 
This type of character can only be obtained by modified general relativity or exotic matter.

\item 
We have determined that $f(T)$ is the uniformly expanding function in our model (see Fig. $5$).

\item 
The EoS parameter $\omega_{D}$ is a crucial parameter describing the various matter-dominated phases of the universe's evolution. 
The overall effect of the EoS parameter demonstrates that the universe began at $\omega_{D}= 0$ (goes up to negative values and represents 
the universe's acceleration), then gradually increases to a positive deceleration phase. Subsequently, it proceeds to the accelerated era, 
the second phase of the universe's current acceleration. As a result, our derived model is found in both the quintessence and phantom regions 
(see Fig. $6$).

 \item 
We have investigated the evolution of dark energy by observing the behaviour of $\Omega_{D}$. We noticed that $\Omega_{D}$ is positive decreasing 
functions (see Fig. 7).

\item 
We have shown how the energy conditions (ECs) are changing with time. It is important to mention here that the SEC must violate to support 
the universe's late-time acceleration, and Fig. 8 shows that SEC is not satisfied.

\item 
Additionally, the $\omega_{D}-\omega_{D}'$ plane revealed that the Renyi HDE model lies in the freezing region for various values of $\delta$ 
during the evolution (see Fig. $9$). In the framework of $f(T)$ modified gravity, the cosmic expansion will be more accelerated. 

\end{itemize}
As a result, in the framework of $f(T)$ gravity, general relativity results are consistent with the Renyi HDE model.
We note that the dark energy issue may be simply represented in Teleparallel gravity as a geometric term.
The scientists who are engaged in this subject are intrigued by this fact. 

\section*{Acknowledgments} 
A. Pradhan also expresses gratitude to the IUCAA, Pune, India for offering facilities and assistance through the visiting associateship programme. 


\end{document}